\documentclass[10pt,osajnl2,letterpaper,twocolumn,showpacs]{revtex4}  

\usepackage{graphicx} 
\usepackage{amsmath}   
\usepackage{amssymb}   
\usepackage{color}

\begin{document}

\newcommand {\edit}[1]{\textcolor{red}{#1}}

\title{A crossed vortex bottle beam trap for single-atom qubits}

\author{G. Li, S. Zhang, L. Isenhower, K. Maller, and M. Saffman$^*$}

\affiliation{
Department of Physics, University of Wisconsin, 1150 University Avenue, Madison, Wisconsin 53706, USA\\
$^*$corresponding author: msaffman@wisc.edu
}

\begin{abstract}We demonstrate trapping and quantum state control of single Cesium atoms in a 532 nm wavelength bottle beam trap. The three dimensional trap is formed by crossing two unit charge vortex beams. Single atoms are loaded with 50\% probability directly from a magneto-optical trap. We achieve a trapping lifetime of up to 6 s, and demonstrate fast Rabi oscillations with a coherence time of $T_2\sim 43 \pm 9\rm~ ms$.\end{abstract}

\date{\today}


\maketitle 

Qubits encoded in the hyperfine states of single neutral atoms that are confined in an array of optical traps represent
a promising and actively pursued approach to implementing multi-qubit quantum 
information processing (QIP) devices\cite{Meschede2006,Wilk2010,Isenhower2010,Lengwenus2010}.  
Far detuned optical traps provide strong confinement with low photon scattering rates and low decoherence.
It has been possible to load single atoms into micron sized traps with $\sim 50\%$ probability using collisional blockade\cite{Schlosser2001}
and up to 83\% probability with repulsive light assisted collisions\cite{Grunzweig2010}. Experiments with a few optical traps spaced
by several microns allow for site specific quantum state control and measurements which has led to recent demonstrations of a two-atom CNOT gate\cite{Isenhower2010} and entanglement\cite{Wilk2010,Zhang2010}. An alternative to arrays of optical dipole traps is to use optical lattices with a sub-micron separation between trap sites. The BEC-Mott insulator transition can be used for close to unity loading of these short period arrays\cite{Greiner2002},
and recent experiments have demonstrated site resolved imaging\cite{Bakr2009}, as well as quantum state control 
of individual atoms\cite{Weitenberg2011}.

QIP experiments based on Rydberg state mediated interactions of neutral atoms\cite{Saffman2010} present special requirements for the optical trap potentials. Blue detuned traps which hold atoms at a local minimum of the intensity are preferable in order to minimize photoionization of Rydberg states and to equalize the ground and Rydberg state  trapping potentials\cite{SZhang2011}. In addition the trap size, and therefore the lattice spacing, should be large enough to accommodate the wavefunction of the Rydberg electron which has a diameter $> 1.5~\mu\rm m$ for principal quantum number $n\sim 100$. These considerations point towards trap arrays, or long period lattices\cite{Nelson2007}, as promising approaches for experiments using Rydberg atoms. In this letter we report on trapping and quantum state control of single Cs atoms in a far off resonance bottle beam trap (BBT) using 532 nm trapping light which is detuned from the strong 
 Cs $6s_{1/2} - 6p_{3/2}$  transition by 210 THz. This is the largest detuning of any blue detuned optical trap demonstrated to date  which helps to minimize motional qubit decoherence\cite{Kuhr2005} in the trap.

\begin{figure}[!t]
\centerline{\includegraphics[width=8.8cm]{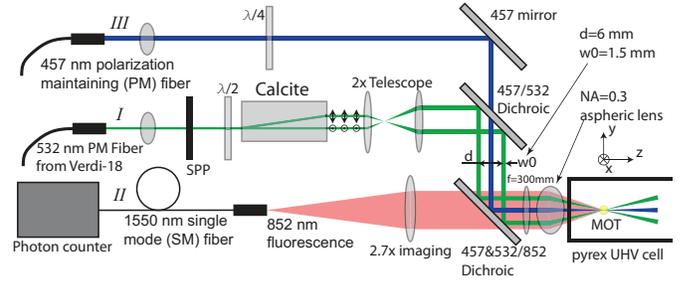}}
\vspace{-.3cm}
\caption{(Color online) Experimental setup with paths for I) 532 nm trap light, II)  852 nm atom observation and III) 457 nm Raman light. }
\label{fig.layout}
\end{figure}

Several previous experiments have demonstrated trapping of many atoms in blue detuned dipole traps\cite{Kuga1997,Ozeri1999,Kulin2001,Olson2007,Isenhower2009} and  single atom trapping was demonstrated in \cite{Xu2010}.   
Here we use an adaptation of the method of \cite{Fatemi2007} which is based on crossing two optical vortex beams to create a three dimensional BBT. 
As shown in Fig. \ref{fig.layout} a vortex beam is produced by sending a single frequency 532 nm beam through a spiral phase plate (SPP) etched in fused silica. We have designed the  optical system using a Laguerre-Gauss approximation to the vortex beam, although the actual transverse structure and propagation dynamics are more complex\cite{Mawardi2011}. A 27 mm long calcite beam displacer separates the beam into orthogonally polarized parallel beams which 
are then focused into a pyrex cell UHV vacuum chamber using a custom designed aspheric lens ($NA=0.3$ and $f=24~\rm mm$).  The aspheric lens  is corrected for the 1.5 mm thick Pyrex vacuum window by adding a 300 mm lens before it.

The vortex beams intersect at the lens focus which coincides with the waist position of each beam. The transverse ($x-y$ plane) light intensity distribution in the focal region is checked by a microscope with resolution below 2$\mu \rm m$ and the images are recorded by a CCD camera. Figure 2(a) shows intensity images at several $z$ positions. From these images the intensity distributions in the $x-z$ and $y-z$ planes are reconstructed (Fig. 2 b and c).  From Fig. 2 we see that the trap has a size of about $22~\mu \rm m$ along $z$ and $3.3~\mu \rm m$ along x and y directions. Calculations predict\cite{SZhang2011} that with $0.48~\rm W$ of 532 nm light the minimum trap barrier is $\sim k_B \times 300~\mu \rm K$, where $k_B$ is the Boltzmann constant.

\begin{figure}[!t]
\centerline{\includegraphics[width=8.5cm]{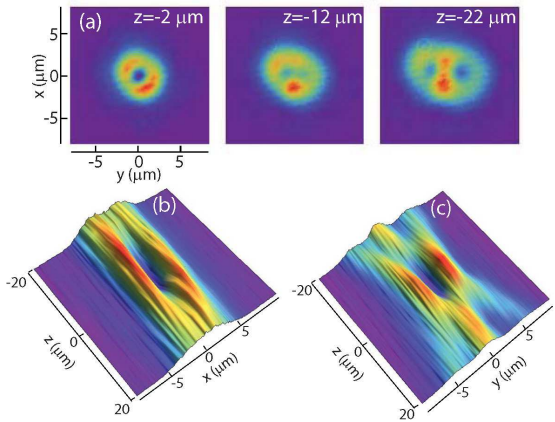}}
\vspace{-.3cm}
\caption{(Color online) Intensity distribution of the BBT recorded by a CCD camera at a) $z=-2, -12, -22 ~\mu \rm m$ and the reconstructed intensity distributions in the b) $x-z$  and c)  $y-z$  planes.}
\end{figure}

Because the BBT has a repulsive  barrier around the trapping region, cold atoms outside the trap with low kinetic energy will not enter and be trapped. In order to load a micron sized BBT, a cold atom sample with high density should be prepared before trap light is applied. In the experiment an acousto-optic   shutter is used to block the BBT beam while a cold Cs cloud is  formed in a magneto-optical trap (MOT). In addition, because we use a typical trap depth of only several hundred $\mu \rm K$, a long trap lifetime requires atom temperatures on the order of a few tens of $\mu \rm K$.  
The single atom loading sequence is 1)  $t_1$ seconds of MOT loading phase with detuning of $\Delta/2\pi = - 10~\rm  MHz$ followed by a $5~\rm ms$ polarization gradient cooling (PGC) phase with $\Delta/2\pi=-30~\rm MHz$ giving an atom temperature of $\sim 20~\mu \rm K$.  2) The BBT light is then unblocked and the PGC beams are kept on for another $5~\rm ms$. 3) All cooling and repump beams are switched off for $20~\rm ms$ to allow atoms outside the BBT to fall away. 4) Finally, the MOT beams with PGC settings are turned on again as readout light for $t_2$ seconds. Fluorescence from trapped atoms is collected by the atom observation optics shown in Fig. 1 and detected by a single photon counter with $100~\rm ms$ integration time. 

Figure 3a shows the  photon counts in a continuous loading mode with $t_1=t_2=4~\rm s$. Figure 3b shows a histogram of photon counts for 2200 loading cycles, from which a $52.6\%$ single atom loading rate is obtained. We do not have a definitive explanation for why the rate is $>50\%$.  From these plots, two photon count levels corresponding to $0$ and $1$ atoms  can be clearly identified.  No two atom events were observed, even with a shorter  photon counter integration time of $20~\rm m s$.

\begin{figure}[!t]
\centerline{\includegraphics[width=8.4cm]{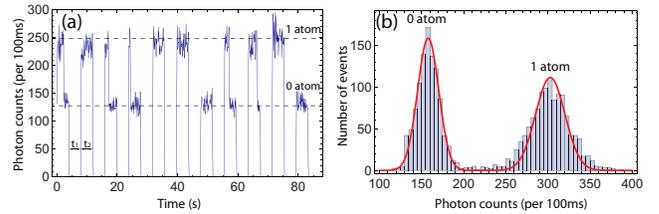}}
\vspace{-.4cm}
\caption{(Color online) Photon counter signal (a) during continuous atom loading process in a $k_B \times 300\mu K$ trap with $t_1,t_2=4~\rm s$. (b) atom loading histogram.}
\end{figure}

Trapped atoms have a finite lifetime due to heating by Raman scattering of the trap light, intensity and pointing noise of the trap light and collisions with  background hot atoms. Figure 4 shows measurements of atom retention in a $k_B\times 300\mu \rm K$ trap under conditions when the readout light is always on (bright trap) and off (dark trap). From the exponential fits we extract $1/e$ lifetimes of  3.8(6.0) s for bright(dark) traps. The longer lifetime of the dark trap, even at large trap depths, is consistent with calculations that account for the $\sim1.5\%$ excited state fraction in the bright trap and larger collisional cross section of optically excited atoms with background Cs\cite{Bjorkholm1988}.

\begin{figure}[!t]
\centerline{\includegraphics[width=8.4cm]{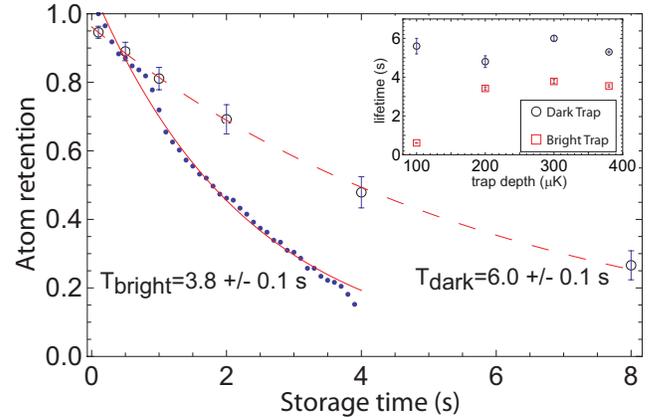}}\vspace{-.4cm}
\caption{(Color online) Atom retention measurements for a single atom in the BBT with $k_B \times 300 \mu \rm K$ trapping depth when readout light is always on (bright trap, small blue dots) and off during storage time (dark trap, open black circles). Each data point is averaged over more than 100 samples. The inset shows the dependence on trap depth.}
\end{figure}

Single atom qubits can be encoded in the Cs clock states $|0\rangle\equiv|f=3,m=0\rangle$ and $|1\rangle\equiv|f=4,m=0\rangle$.
A critical parameter for QIP applications is the coherence time of the  qubits.  
We prepare single atoms in $|1\rangle$ using  randomly polarized repumper light from $6s_{1/2},f=3\rightarrow 6p_{3/2},f=4$ and a $z$ polarized 894 nm beam coupling $6s_{1/2},f=4\rightarrow 6p_{1/2},f=4$. Using  a two-frequency Raman laser system at 457 nm detuned by $\Delta/2\pi = 40~\rm GHz$ from the $6s_{1/2}\rightarrow 7p_{3/2}$ transition we drive single qubit rotations between $|0\rangle$ and $|1\rangle$ at Rabi frequencies up to 1 MHz. With a standard Ramsey sequence ($\pi/2$ pulse, wait for $t_d$, $\pi/2$ pulse) we measured the $T_2$ coherence time of the trapped atoms as shown in Fig. 5. 

The $T_2$ time is primarily limited by motional decoherence due to the differential trap shift of the qubit states and by magnetic noise giving a quadratic Zeeman shift. Since $T_2$ scales inversely with the atomic temperature\cite{Kuhr2005} we further cooled the atoms to about  $4~\mu\rm K$ by applying a $5~\rm ms$ PGC phase at $\Delta/2\pi=-50~\rm MHz$ before optical pumping. For 532 nm trap light the calculated coherence time due to motional decoherence alone at our measured temperature of $4~\mu\rm K$ is $T_2=88~\rm ms$. The observed $T_2=43~\rm ms$ can be explained by $\sim1~\mu\rm T$ of magnetic field noise at our bias field along $z$ of $0.15~\rm mT$. This $T_2$ time compares favorably with recent single or few atom experiments which reported coherence times as long as 20 ms\cite{Kuhr2005} without applying echo pulses. We anticipate that echo pulse sequences and compensation of the differential hyperfine shift caused by the trap light\cite{Radnaev2010} will significantly improve the $T_2$ reported here.

\begin{figure}[!t]
\centerline{\includegraphics[width=8.1cm]{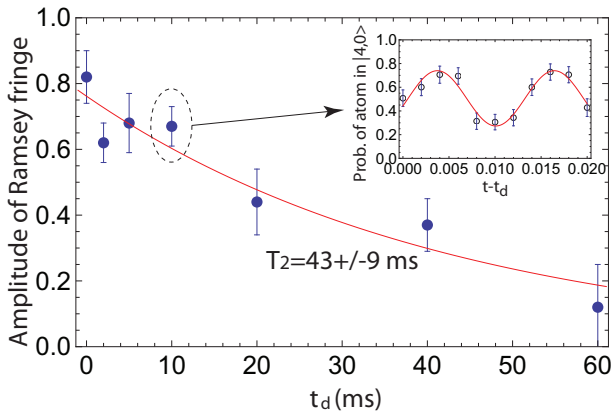}}
\vspace{-.45cm}
\caption{(Color online) Ramsey fringe contrast (blue dots) at different delay times $t_d$ between $\pi /2$ Raman pulses. The exponential fit (red curve) gives $T_2=43\pm9~\rm ms$. The non-unity contrast at $t_d=0$ is due to noise on the Raman lasers. Inset: Ramsey fringe measurement at $t_d=10 ~\rm ms$. Eeach data point comes from an average of more than 40 samples.}
\end{figure}

In summary, we have demonstrated a $\mu\rm m$ sized crossed-vortex BBT  and shown that single Cs atoms can be loaded with 50 \% probability. With atom lifetimes of several seconds and coherence times of $\sim 43~\rm ms$
the BBT is a promising building block for multi-qubit experiments. 
The BBT geometry has the potential for trapping atoms in Rydberg states\cite{SZhang2011,Younge2010} and is therefore attractive for Rydberg mediated QIP experiments. Using diffractive optical beam splitters we have demonstrated 2D arrays of BBTs and are currently investigating atom loading into multiple sites.

The work was supported by the IARPA MQCO program through ARO contract W911NF-10-1-0347 and DARPA.


\newpage

\section*{Informational Fourth Page}

\end{document}